\documentstyle[12pt,psfig]{article}
\let\section=\subsection     \let\subsection=\subsubsection                

\begin{document}
\newcommand \beq{\begin{eqnarray}}
\newcommand \eeq{\end{eqnarray}}
\newcommand \ga{\raisebox{-.5ex}{$\stackrel{>}{\sim}$}}
\newcommand \la{\raisebox{-.5ex}{$\stackrel{<}{\sim}$}}
\begin{center}
   {\large \bf Kaons in Nuclear Matter}\\[5mm]
   Henning Heiselberg \\[5mm]
   {\small \it  NORDITA \\
   Blegdamsvej 17, DK-2100 Copenhagen \O, Denmark \\[8mm] }
\end{center}

\begin{abstract}\noindent
The kaon energy in a nuclear medium and its dependence on kaon-nucleon
and nucleon-nucleon correlations is discussed.  The transition from the
Lenz potential at low densities to the Hartree potential at high
densities can be calculated
analytically by making a Wigner-Seitz cell approximation and employing
a square well potential. 
As the Hartree potential is less attractive than the Lenz one, kaon
condensation inside cores of neutron stars appears to be less likely than
previously estimated.
\end{abstract}

\section{Introduction}
 Kaon condensation in dense matter was suggested by Kaplan and Nelson
\cite{KN}, and has been discussed in many recent publications
\cite{BLRT,Weise}. Due to the attraction between $K^-$ and
nucleons its energy decreases with increasing density, and eventually
if it drops below the electron chemical potential in neutron star matter in
$\beta$-equilibrium, a Bose condensate of $K^-$ will appear.
It is found that $K^-$'s condense at densities above
$\sim 3-4\rho_0$, where $\rho_0=0.16$ fm$^{-3}$ is normal nuclear matter
density. This is to be compared to the central density of
$\sim4\rho_0$ for a neutron star of mass 1.4$M_\odot$ according to the
estimates of Wiringa, Fiks and Fabrocini \cite{WFF} using realistic
models of nuclear forces. The condensate could change the structure
and affect maximum masses and cooling rates of neutron stars
significantly.

Recently, \cite{PPT,kaon} we have found that the kaon-nucleon and
nucleon-nucleon correlations conspire to reduce the $K^-N$ attraction
significantly already at rather low densities when the interparticle
distance is comparable to the range of the $KN$ interaction.  We have
calculated the kaon energy as function of density in a simple model
where also the low and high density limits and the dependence on the
range of the interaction can be extracted.

\section{Low Density Limit: Lenz Potential}
In neutron matter at low densities when the interparticle spacing is much
larger than the range of the interaction, $r_0\gg R$,
the kaon interacts strongly many times with the same nucleon before it
encounters and interacts with another nucleon. 
Thus one can use the scattering length as the ``effective'' kaon-nucleon
interaction, $a_{K^-n}\simeq -0.41$fm. The kaon energy deviates from its 
rest mass by the Lenz potential
\beq
   \omega_{Lenz} = m_K + \frac{2\pi}{m_R}\, a_{K^-n}\,\rho
               \,.\label{Lenz}
\eeq
which is the optical potential obtained in the impulse 
approximation.

\section{High Density Limit: Hartree Potential}
At high densities when the interparticle spacing is much
less than the range of the interaction, $r_0\ll R$, the kaon
will interact with many nucleons on a distance scale much less the range
of the interaction. 
The kaon thus experiences the field from many nucleons
and the kaon energy deviates from its rest mass 
by the Hartree potential:
\beq
   \omega_{Hartree} = m_K + \rho \int V_{K^-n}(r)
           d^3r          \,,\label{Hartree}
\eeq
As shown in \cite{PPT}, the Hartree potential is considerably less attractive
than the Lenz potential. This is also evident from Fig. (1).

\section{General Case}
To demonstrate this transition from the low density Lenz potential to
the high density Hartree potential we solve the Klein-Gordon equation
for kaons in neutron matter in the Wigner-Seitz cell
approximation. With the simplified square well potential 
this can in fact be done analytically.

\subsection{The Recoil Corrected Klein-Gordon Equation}
We choose to describe the kaon-nucleon interaction by a vector potential
$V$ dominated by the Weinberg-Tomozawa term. 
In the analysis of Ref. \cite{Speth} the $K^+N$ interaction was also
found to be dominated by $\omega$ and $\rho$ vector mesons.
The energy of the kaon-nucleon center-of-mass system with respect to
the nucleon mass is then
\beq
   \omega = \sqrt{k^2+m_K^2}+V+\frac{k^2}{2m_N}  \,, \label{E}
\eeq
where $k$ is the kaon momentum (we use units such that $\hbar=c=1$), 
in c.m. frame.
We have included the recoil kinetic energy of the nucleon assuming that 
terms of order $k^4/8m_N^3$ and higher can be neglected.
For a relativistic description of the kaon in a vector potential we
employ the following recoil corrected Klein-Gordon equation (RCKG) 
obtained by quantizing Eq. (\ref{E}) ($k=-i\nabla$)
\beq
   \left\{ (\omega-V)^2 +\frac{m_N+\omega-V}{m_N}\nabla^2-m_K^2 \right\} 
     \phi = 0       \,.\label{RCKG}
\eeq

\subsection{The Wigner-Seitz Cell Approximation}
The Wigner-Seitz cell approximation simplifies band structure
calculations enormously. Though it is a poor approximation for solids
it is better for liquids. As we only consider qualitative effects we
shall assume the Wigner-Seitz cell approximation for the strongly
correlated nuclear liquid because the periodic boundary condition is a
computational convenience. It contains the important scale for
nucleon-nucleon correlations given by the interparticle spacing and,
as we shall see, it naturally gives the correct low density (Lenz) and
high density (Hartree) limits. We only consider neutrons since the
$\sim10\%$ protons expected in neutron stars do not change results by
much.

\subsection{Square Well Potential}
Since the kaon-nucleon potential is not known in detail
and we are only interested in qualitative effects, we will for simplicity
approximate it by a simple square well potential
\beq
   V(r)= -V_0\Theta(R-r) \,.\label{VSW}
\eeq
The range of the interaction $R$ and the potential depth $V_0$ are
related through the s-wave scattering length
$a=R-\tan(\kappa_0R)/\kappa_0$, where
$\kappa_0^2=(2m_KV_0+V_0^2)\cdot$ $m_N/(m_N+m_K+V_0)$.
If $V_0\ll m_K$ this reduces to the standard result 
$\kappa_0^2=2m_RV_0$, where
$m_R=m_Km_N/(m_K+m_N)$ is the kaon-nucleon reduced mass. 
However, for small ranges of interaction
the potential becomes significant as compared to
the kaon mass which necessitates a relativistic treatment 
with the Klein-Gordon equation instead of
the Schr\"odinger equation.

The above arguments suggest that the range
$R = 0.4-1.0$fm covers most realistic possibilities above.
For a kaon-neutron scattering length of $a_{K^-n}=-0.41$ fm we find for
typical ranges $R=0.4-1.0$ fm kaon-neutron potentials  ranging from
$V_0=463$MeV down to $V_0=49$ MeV.

\noindent
\begin{minipage}[b]{16cm}
{\centering
\mbox{\psfig{file=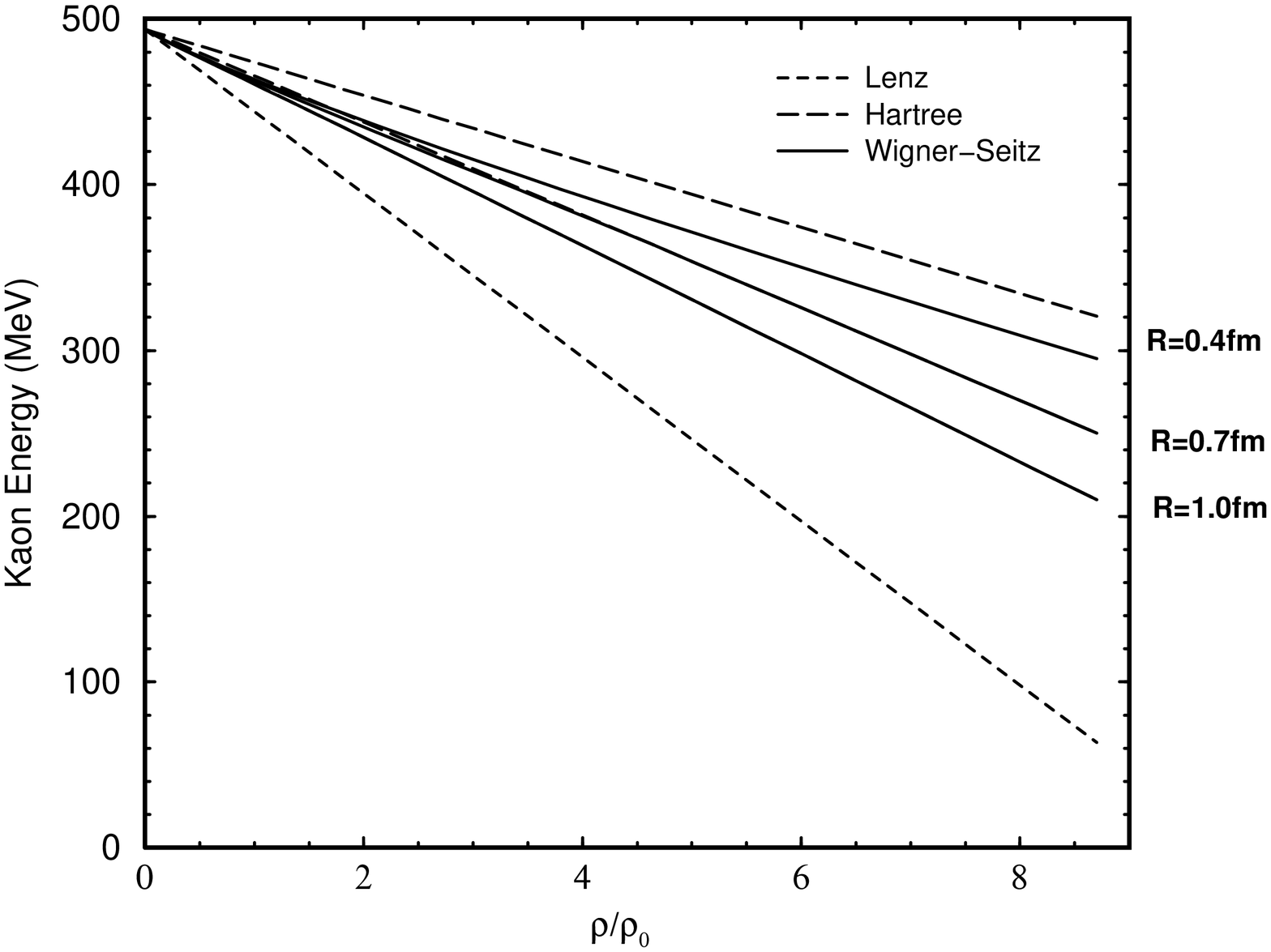,height=100mm,angle=-90}}}
\end{minipage}
\vspace*{-5mm}
\begin{center}
\begin{minipage}{13cm}
\baselineskip=12pt
{\begin{small}
Fig.~1. 
Kaon energy as function of neutron density. Our calculation
(Eq. (\ref{3}), full curves) are shown for $R=0.4$ fm, $R=0.7$ fm and
$R=1.0$ fm. At low densities they approach the Lenz result
(Eq. (\ref{Lenz}), dotted curve) and at high densities they
approach the Hartree result (Eq.(\ref{Hartree}), dashed curves).
\end{small}}
\end{minipage}
\end{center}

\section{Results}
As shown in \cite{kaon} 
the RCKG equation can now be solved for s-waves and the kaon wave-function
found.
Applying the periodic boundary condition, $\phi'(r_0)=0$, as
required by the Wigner-Seitz cell approximation we arrive at
the closed equation
\beq
   \frac{k}{\kappa}\tan(\kappa R)=\frac{e^{2k(R-r_0)}-(1-kr_0)/(1+kr_0)}
         {e^{2k(R-r_0)}+(1-kr_0)/(1+kr_0)} \,, \label{3}
\eeq 
which determines $k$ and thus the kaon energy. 
Here,
$k^2=(m_K^2-\omega^2)m_N/(m_N+\omega)$ and  
$\kappa^2=\left((\omega+V_0)^2-m_K^2\right)m_N/(m_N+\omega+V_0)$.
The resulting kaon energy is
shown in Fig. (1). By expanding (\ref{3}) at low densities the kaon
energy becomes the Lenz potential whereas at high densities
it becomes the Hartree potential.

\section{Summary}

Kaon-nucleon correlations reduce the $K^-N$
interaction significantly when its range is comparable to or larger
than the nucleon-nucleon interparticle spacing.  The transition from
the Lenz potential at low densities to the Hartree potential at high
densities begins to occur already well below nuclear matter densities.
For the measured $K^-n$ scattering lengths and reasonable ranges of
interactions the attraction is reduced by about a factor of 2-3 in
cores of neutron stars. Relativistic effects further reduce the
attraction at high densities.  Consequently, a kaon condensate is less
likely in neutron stars. 

Coulomb energies have not been included in the above analysis. They may,
as discussed in \cite{Schaffner,rotp}, lead to a mixed phase of nuclear
matter with and without a kaon condensate. However, the Coulomb energies
are small as compared to kaon masses and therefore
the reduction in the kaon energy will also be minor.

My collaborators J.Carlson, V.Pandharipande and C.J.Pethick are
gratefully acknowledged.

\end{document}